\documentclass[11pt]{article}
\usepackage[english]{babel}
\usepackage[utf8]{inputenc}
\usepackage{fancyhdr}
\usepackage{authblk}
\usepackage{xspace}
\usepackage{amsfonts}
\usepackage{amsmath}
\usepackage{times} 
\usepackage{graphicx}
\usepackage[left=1.2in,right=1.2in,top=3cm,bottom=3cm]{geometry} 
\usepackage{multirow}
\usepackage[table,xcdraw]{xcolor}
\usepackage{array}
\usepackage{relsize}
\usepackage{tabu}
\usepackage{url}
\usepackage{listings}
\usepackage{natbib}
\usepackage[labelfont=bf]{caption}

\usepackage[inline,shortlabels]{enumitem}
\definecolor{codegreen}{rgb}{0,0.6,0}
\definecolor{codegray}{rgb}{0.5,0.5,0.5}
\definecolor{codepurple}{rgb}{0.58,0,0.82}
\definecolor{backcolour}{rgb}{0.95,0.95,0.92}

\lstdefinestyle{mystyle}{
    backgroundcolor=\color{backcolour},   
    commentstyle=\color{codegreen},
    keywordstyle=\color{magenta},
    numberstyle=\tiny\color{codegray},
    stringstyle=\color{codepurple},
    basicstyle=\ttfamily\footnotesize,
    breakatwhitespace=false,         
    breaklines=true,                 
    captionpos=b,                    
    keepspaces=true,                 
    numbers=left,                    
    numbersep=5pt,                  
    showspaces=false,                
    showstringspaces=false,
    showtabs=false,                  
    tabsize=2
}
\usepackage{setspace}

\title{{Distributional regression models for 
Extended Generalized Pareto distributions}}
\author[1]{No\'emie Le Carrer\footnote{Corresponding author: noemie.le-carrer@imt-atlantique.fr}}
\affil[1]{Lab-STICC, IMT Atlantique  Brest, France}
\author[2]{Carlo Gaetan}
\affil[2]{Dipartimento di Scienze Ambientali, Informatica e Statistica \newline
	Universit\`a Ca' Foscari  Venezia, Venice, Italy}

\begin{document}

\maketitle
\renewenvironment{abstract}
 {\quotation\small\noindent\rule{\linewidth}{.5pt}\par\smallskip
  {\centering\bfseries\abstractname\par}\medskip}
 {\par\noindent\rule{\linewidth}{.5pt}\endquotation}

\begin{abstract}
The Extended Generalized Pareto Distribution (EGPD) (Naveau et al. 2016) is a family of distribution that has been introduced to model the full range of a positive random variable but with the lower and the upper tails distributed according the peaks-over-threshold methodology. The aim of this article is to augment the scope of application of EGPD allowing the analyst to incorporate the effect of covariates on the model. In particular we introduce a specification where the parameters of EGPD can be modeled as additive functions of the covariates, e.g. space or time. As a related product we provide an add-on code written in R that it is flexible enough to implement the EGPD in a generic way, allowing to
introduce new parametric forms. We show the potential of our add-on on  the modeling of hourly rainfalls over the North-West region of France and discuss modeling strategies.   
\end{abstract}


\newpage

\section{Introduction}
\label{secIntro}

The Peak Over Threshold (POT) method, described in details by \citet[][Chapter 4]{coles2001introduction},  
has been used in many fields to identify extremal events such as loads,
wave heights, floods, wind velocities, etc. This method provides a
model for independent exceedances over a high threshold.
According the  
Pickands-Balkema-de Haan theorem \citep{balkema:dehaan:1974,pickands:1975}
for a threshold $u$  sufficiently high, the unnormalized conditional tail distribution of a random variable $Y$ can be approximated as 
$$
\Pr(Y-u\le y|Y>u)\approx 	H_{\xi}(y/\psi_u),\qquad \text{for}\, y\ge 0\quad \text{and}\quad \psi_u >0
$$
where
\begin{equation}\label{eq:GPD}
	H_{\xi}(z) =
	\begin{cases}
		1 - {(1+\xi y)}_+^{-1/\xi} & \text{for $\xi \neq 0$}\\
		1-\exp\{-y \}& \text{for $\xi = 0$}
	\end{cases}  
\end{equation}
and  $a_+ = \max (a, 0)$. 

 The limiting distribution \eqref{eq:GPD} is referred to as the Generalized Pareto  distribution (GPD) and the parameters $\xi$ and $\psi_u$ are respectively the shape and the scale parameters. 
 The scale parameter depend on the threshold $u$, but 
in  remainder of the paper the subscript $u$ is  dropped in the  notation.

Modeling using GPD has become very popular since the paper of \citet{davison:smith:1990} where estimation and model-checking procedures for univariate and regression data are developed.
 Such procedure have then been extended  by allowing the GPD parameters to be represented as smooth functions of covariates \citep{chavez-demoulin:davison:2005}.
 
 Since then several R packages have been developed to fit GPDs  and contributed to the Comprehensive R Archive Network (CRAN) in the field of extreme value analysis \citep{team2013r}. Among them, \texttt{ismev} \citep{heffernan2016ismev}, providing tools for the  extreme value analyses presented in \citet{coles2001introduction}, \texttt{evd} \citep{stephenson2002evd}, \texttt{evir} \citep{pfaff2018package}, \texttt{extRemes} \citep{gilleland2016extremes} and \texttt{mev} \citep{belzile1mev}, have offered various approaches for fitting univariate and multivariate GPDs. \citet{gilleland2013software} and, more recently, \citet{Belzile-et-al:2022} review them while the \citet{dutang2022cran}'s CRAN Task View provides an up-to-date list of the R packages in this field. In particular, some of them offer regression-based models for extreme distributions, where GPD parameters are allowed to vary with covariates. 
Beyond the linear forms introduced in the packages \texttt{ismev} and \texttt{evd},  
a few R packages allow to fit GPDs with parameters following additive forms i.e. when the parameters or their transformations can be modelled as additive smooth functions of the covariates. 
In the sequel we term this form a Generalized Additive Model (GAM) \citep{Hastie:Tibshirani:1990,wood2017generalized}. Thus \texttt{VGAM} \citep{yee2007vector}, \texttt{gamlss} \citep{rigby2005generalized} and \texttt{evgam} \citep{youngman2020evgam} provide this functionality for GPDs, as reviewed among other alternatives in \citet{youngman2020evgam}.

It is clear that GPD  needs
the determination of a threshold which is neither too high.
The choice of an appropriate threshold is  challenging because the stability plots and the mean residual life
\citep[][Chapter 4]{coles2001introduction} plot do not give an obvious value of a suitable threshold. 
The problem is tackled  in a regression setting by
\citet{eastoe:tawn:2009}. 
However the question of the threshold choice and then simulation of the whole spectrum of the random variable considered remains problematic. 

Forming mixtures of Exponential distributions \citep{wilks1998multisite,keller2015implementation} or conditional Gamma distributions \citep{kenabatho2012stochastic,chen2014multi} remains a competitive alternative e.g. for heavy-tailed precipitation modeling and simulation. Approaches of this type are implemented in the package \texttt{GSM}, that uses a Bayesian approach for estimating a mixture of Gamma distributions in which the mixing occurs over the shape parameter and the procedure only requires to specify a prior distribution for a single parameter. The package \texttt{evmix} fits an extreme value mixture model with normal behavior for the bulk distribution up to the threshold and conditional GPD above the threshold. Options for profile likelihood estimation for threshold and fixed threshold approach are provided. Finally, \texttt{distributionsrd} provides the double Pareto-lognormal distribution.

Although interesting, in practice these packages do not allow to fit nonlinear models as flexible as the GAM options allowed for the GPD alone.
Besides, the mixture approaches tend to quickly inflate the number of parameters and the inference of the latter may be problematical operationally \citep{naveau2016modeling}.

\citet{papastathopoulos2013extended} opened another route in the field of heavy-tail distribution modeling by replacing the uniform draws in the inverse cumulative distribution function (CDF) transform to generate simulations of the GPD, by draws from a richer distribution e.g. type Beta distribution. Using this approach, \citet{naveau2016modeling} presented a concise parametric formulation (later extended to a semi-parametric formulation \citep{tencaliec2020flexible}) for simulating the whole range of a positive random variable (namely precipitations), that makes use of GPD to model not only the high quantiles but the smallest ones as well and bypass the issue of threshold selection. 
The main idea rooted from composing a cumulative distribution function with the GPD (see section 2 for more details). Very recently, a similar idea was taken up by \citet{stein2021env,stein2021ext}.

In the sequel, we refer to the extension of GPD in the sense of \citet{naveau2016modeling} as Extended Generalized Pareto distribution (EGPD).
An implementation of EGPD with four parametric families, together with the fitting procedure, can be found  in the R package \texttt{mev}.

The aim of this work is to extend the applicability of the EGPD in the presence of covariates that can modify the parameters of the distribution.
The extension takes place in the same spirit as  GAM. 
We also want to present an add-on to the \texttt{gamlss} package that allows to make inference on the parameter.

We have chosen the \texttt{gamlss} package because it provides univariate distributional regression models, where parameters of the assumed distribution for the response can be modeled as additive functions of the covariates. It is flexible enough to allow the encoding of novel families of distribution, as long as their density, cumulative, quantile and random generative function and, optionally, first and second (cross-) derivatives of the likelihood can be computed.  We seize this potential to implement the parametric EGPD family in a generic way, allowing to introduce new parametric forms, up to 4 parameters so far (due to limitations of \texttt{gamlss}), satisfying the requirement of the EGPD family.

The remaining of this article will first introduce the EGPD and the  distribution  regression model (Section \ref{secTheory}), followed by the key commands to use the EGPD within the \texttt{gamlss} package and our add-on in Section \ref{secTuto}. 
Section \ref{secApplRainfalls} implements them to perform an analysis of hourly positive rainfalls on the North-west of France and discuss specific advantages and limitations of distributional regression modeling and its implementation in R, while Section \ref{secConclu} concludes the paper. 

\section{A distributional regression model}
\label{secTheory}

\subsection{Extended-GPD}
\label{secEGPDmod}

The EGPD \citep{naveau2016modeling}  accounts simultaneously and in a parsimonious way for both extreme and non extreme values of a random variable $Y$. Its cumulative distribution function (CDF) reads:
\begin{equation}
\label{CDFegpd}
    F(y) = G\{ H_{\xi}(y/\psi) \} \:, \: \: \text{for all}\quad y > 0.
\end{equation}
The function $G(\cdot)$ is a continuous CDF on the unit interval. It is constrained by several assumptions: 1) we need to ensure that the upper-tail behavior of $F$ follows a GPD; 2) the low values of $Y$, seen as the upper tail of $-Y$ when $y \rightarrow 0$, follows a GPD as well. 

This leads to the following consequences for any candidate CDF $G(\cdot)$: 
\begin{enumerate}
	\item[a.]  the upper-tail behavior
	of $1-F(y)$ is equivalent to the original GPD tail used to build $F(y)$ ; 
	\item[b.]  when $y \rightarrow 0$, $\displaystyle F(y) \sim \frac{c}{\psi^s}y^s $ where $\displaystyle c = \lim_{y \rightarrow 0} \frac{G(y)}{y^s}$ is positive and finite for some real $s$.
\end{enumerate}

Four parametric families, $G(\cdot,\kappa)$, $\kappa=(\kappa_1,\ldots,\kappa_k)^T$, that satisfies the above-mentioned constraints are proposed in \citet{naveau2016modeling}.

\begin{enumerate}
    \item \label{c1} 
    $G(u;\kappa)=u^{\kappa_1}$, $\kappa_1 >0$
    \item \label{c2}
    $G(u;\kappa)=\kappa_3 u^{\kappa_1}+(1-\kappa_3)u^{\kappa_2}$, $\kappa_2 \geq \kappa_1 >0$ and $\kappa_3 \in [0,1]$
    \item \label{c3}
    $G(u;\kappa)=1-Q_{\kappa_1}\{(1-u)^{\kappa_1}\}$, $\kappa_1>0$ where $Q_{\kappa_1}$ is the CDF of a Beta random variable with parameters $1/\kappa_1$ and $2$, that is:
    $$Q_{\kappa_1}(u)=\frac{1+\kappa_1}{\kappa_1} u^{1/\kappa_1} \Big( 1-\frac{u}{1+\kappa_1} \Big)$$
    \item \label{c4}
    $G(u;\kappa)=[1-Q_{\kappa_1}\{(1-u)^{\kappa_1}\}]^{\kappa_2/2}$, $\kappa_1, \kappa_2>0$
\end{enumerate}

In model \ref{c1}, $\xi$ controls the rate of upper tail decay, $\kappa_1$ the shape of the lower tail and $\psi$ is a scale parameter. $\kappa_1=1$ allows us to recover the GPD model, while varying it gives flexibility in the description of the lower tail. 

Model \ref{c2}, as a mixture of power laws, allows to increase the flexibility provided by model \ref{c1}. Therein, $\kappa_1$ controls the lower tail behavior while $\kappa_2$ modifies the shape of the density in its central part. 

Model \ref{c3} is connected to the work of \citet{falk2010laws}. $\xi$ keeps controlling the upper extreme tail, $\kappa_1$ the central part of the distribution in a threshold tuning way. However, this model imposes a behavior type $\chi^2$ for the lower tail (null density at $0$) instead of allowing the data to constrain it. 

Hence Model \ref{c4}, which adds an extra parameter $\kappa_2$ to circumvent this limitation. In this last model, $\kappa_2, \kappa_1, \xi$ control respectively the lower, moderate and upper parts of the distribution. By construction, the lower and upper tails are GPD of shape parameters $\kappa_2$ and $\xi$ respectively.

In the sequel we denote the vector of parameters $(\xi,\psi,\kappa^T)^T$ with $\theta = (\theta_1, ..., \theta_d)^T$ and the density of a parametrized EGPD with $EGPD(y;\theta)$.
To simulate from Equation (\ref{CDFegpd}), the user randomly draws a uniform variable $U$ in $[0,1]$ and then applies the quantile function $Y=F^{-1}(U)$, given for $p \in [0,1]$ by:
\begin{equation}
\label{quantileegpd}
     y_p=F^{-1}(p) =
     \begin{cases}
       \frac{\psi}{\xi}[\{ 1-G^{-1}(p) \}^{-\xi} - 1] & \text{for $\xi \neq 0$}\\
       -\psi \log{ \{ 1-G^{-1}(p) \}} & \text{for $\xi = 0$}
    \end{cases}  
\end{equation}

\subsection{Additive model specification}
\label{GAM}

In practice, the parameters of the distribution of $Y$ may depend on  some covariates $\boldsymbol{x}=(x_1,\ldots,x_m)^T$, i.e.
\begin{equation}\label{eq:EGPD}
Y|\boldsymbol{x}\sim EGPD(\cdot,\theta(\boldsymbol{x}))
\end{equation}
where $\theta(\boldsymbol{x})=(\theta_1(\boldsymbol{x}),\ldots,\theta_d(\boldsymbol{x}))^T$.
The previous specification is an instance  of a  distributional regression model \citep{Stasinopoulos-et-al-2018}. 

For relating the different distributional parameters  $(\theta_1(\boldsymbol{x}),\ldots,\theta_d(\boldsymbol{x}))$ to the covariates, we rely on additive predictors of the form
\begin{equation}\label{eq:predictors}
\eta_i(\boldsymbol{x})=f_{i1}(\boldsymbol{x})+\cdots+f_{iJ_i}(\boldsymbol{x})	
\end{equation}
where $f_{i1}(\cdot),\ldots, f_{iJ_i}(\cdot)$ are smooth functions of the covariates $\boldsymbol{x}$.

The representation \eqref{eq:predictors} is particularly flexible and allows to capture complex dependence patterns between distribution parameters and covariates
$\boldsymbol{x}$. For instance a semiparametric model with  two covariates $(x_1,x_2)$ can be written as
$$
\eta_i(\boldsymbol{x})=\beta_{i1}+ \beta_{i2}x_1+f_{i2}(x_2)
$$
Another example is when we want to model  space-time data observed at site $(s_1,s_2)$ and at time $t$. In this case $\boldsymbol{x}=(s_1,s_2,t)^T$
$$
\eta_i(\boldsymbol{x})=f_{i1}(t)+ f_{i2}(s_1,s_2)
$$

In analogy to Generalized Linear Model the predictors are  linked to the distributional parameters via known monotonic and twice differentiable link functions $h_i(\cdot)$.

In the case of model 1, i.e. $G(u,\kappa)=u^\kappa_1$, common link functions are
$$
\xi(\boldsymbol{x})=\eta_\xi(\boldsymbol{x}) \mbox{ (i.e. the identity function)},\quad \psi(\boldsymbol{x})=\exp(\eta_\psi(\boldsymbol{x})),\quad \kappa_1(\boldsymbol{x})=\exp(\eta_{\kappa_1}(\boldsymbol{x})).
$$

\begin{equation}\label{eq:link}
\theta_i(\boldsymbol{x})=h_i(\eta_i(\boldsymbol{x})), \qquad i=1,\ldots, d
\end{equation}
The functions $f_{ij}$in  \eqref{eq:predictors} are approximated in terms of  basis function expansions

\begin{equation}\label{eq:basis}
	f_{ij}(\boldsymbol{x})=\sum_{k=1}^{K_{ij}} \beta_{ij,k}B_k(\boldsymbol{x}),
\end{equation}
where $B_k(\boldsymbol{x})$ are the basis functions and $\beta_{ij,k}$ denote the corresponding basis coefficients. These basis can be of different types, thoroughly presented e.g. in \citet{stasinopoulos2017flexible} or \citet{wood2017generalized}.

The basis function expansions can be written as 
$$
f_{ij}(\boldsymbol{x})=  \boldsymbol{z}_{ij}(\boldsymbol{x})^T\boldsymbol{\beta}_{ij}
$$
where $\boldsymbol{z}_{ij}(\boldsymbol{x})$ is still a vector of transformed covariates that depends on the basis functions.
and $\boldsymbol{\beta}_{ij}=(\beta_{ij,1},\ldots, \beta_{ij,K_{ij}})^T$ is a parameter vector to be estimated.

To ensure regularization of the functions $f_{ij}(\boldsymbol{x})$  so-called penalty terms are added to the objective log-likelihood function. 
Usually, the penalty for each function $f_{ij}(\boldsymbol{x})$ are quadratic penalty $\lambda \boldsymbol{\beta}_{ij}^T \boldsymbol{G}_{ij}(\boldsymbol{\lambda}_{ij}) \boldsymbol{\beta}_{ij} $ where $\boldsymbol{G}_{ij}(\boldsymbol{\lambda}_{ij})$ is a known semi-definite matrix and the vector  $\boldsymbol{\lambda}_{ij}$ regulates the amount of smoothing
needed for the fit. A special case is when $\boldsymbol{G}_{ij}(\boldsymbol{\lambda}_{ij})=\lambda_{ij}\boldsymbol{G}_{ij}$. Therefore the type and properties of the smoothing functions are controlled by the vectors  $\boldsymbol{z}_{ij}(\boldsymbol{x})$ and the matrices $\boldsymbol{G}_{ij}(\boldsymbol{\lambda}_{ij})$.

The penalized log-likelihood function for the latter models reads:
\begin{equation}\label{eq:penlik}
	l_p = l - \frac{1}{2}\sum_{i=1}^d \sum_{j=1}^{J_i} \boldsymbol{\beta}_{ij}^T \boldsymbol{G}_{ij}(\boldsymbol{\lambda}_{ij}) \boldsymbol{\beta}_{ij}
\end{equation}
where $l$ represents the log-likelihood function for the model \eqref{eq:EGPD}.

\section{Distributional regression  model for EGPD within the R package \texttt{gamlss}}
\label{secTuto}

In this section we introduce an add-on script in R 
that implements the EGPD regression model described in the previous section. 
The script, available in
\texttt{github.com/noemielc/egpd4gamlss}, is flexible enough as it only requires the specification of a parametric CDF $G(\cdot)$.

The Generalized Additive Model for Location, Scale and Shape (GAMLSS) is an example of distributional regression model and the companion package in R \texttt{gamlss} \citep{stasinopoulos2017flexible}  allows a straightforward fitting of multiple built-in and well-known family distributions with parameters that vary flexibly with $\boldsymbol{x}$ as in \eqref{eq:predictors}-\eqref{eq:link}.

An interesting feature is the possibility to extend the family of distributions described in \citet{rigby2019distributions}.
The only limitation is the dimension of the parameter vector $\theta$, $d=4$,
where the four parameters are denoted with $\mu$, $\sigma$, $\nu$ and $\tau$ in \texttt{gamlss}.

Due to the restriction on the number of parameters, our code allows the implementation of parametric distributions for $G$ with only 2 parameters, namely $\kappa=(\kappa_1,\kappa_2)$.
In the following, the parameters $(\xi,\psi,\kappa_1,\kappa_2)$ of the EGPD distribution are identified with $(\mu,\sigma,\nu,\tau)$.

The generic implementation of a EGPD is provided in the file \texttt{GenericEGPDg.R}. The \texttt{MakeEGPD} function assembles a new EGPD family   distribution exploiting an user-built $G(\cdot)$ as a function of $u$ and (at most) two parameters $\nu$ and $\tau$. 
The required  first,  second and  cross derivatives of the density in \eqref{eq:EGPD}, necessary for the fitting procedure in \texttt{gamlss}, are obtained by   the symbolic derivation  using the R package \texttt{Deriv}. Moreover the inverse function $G^{-1}(u)$ is  computed numerically.

In the following, we exemplify the use of \texttt{MakeEGPD} by assembling   Model 1, $G(u;\kappa)=u^{\kappa_1}, \kappa_1 > 0$.

\begin{lstlisting}[language=R]
library(gamlss)
source ("GenericEGPDg.R")
EGPD1Family <- MakeEGPD (function (z,nu) z^nu, Gname = "Model1")
EGPDModel1 <- EGPD1Family()
\end{lstlisting}
The last command is necessary and allows  to create the object EGPDModel1 (name concatenating \texttt{EGPD} and \texttt{Gname}) from the \texttt{EGPD1Family} function.

For given values $\mu_0, \sigma_0$ and $\nu_0$, the R functions of
density, cumulative distribution function, quantile function and random generation are ready to use.
\begin{lstlisting}[language=R]
dEGPDModel1(x,mu=mu0,sigma=sigma0,nu=nu0)
pEGPDModel1(x,mu=mu0,sigma=sigma0,nu=nu0)
qEGPDModel1(u,mu=mu0,sigma=sigma0,nu=nu0)
rEGPDModel1(n,mu=mu0,sigma=sigma0,nu=nu0)
\end{lstlisting}

The reserved names  for the distribution parameters in \texttt{gamlss} are $\mu$ \texttt{mu} ($\mu$), \texttt{sigma} ($\sigma$), \texttt{nu} ($\nu$), and \texttt{tau} ($\tau$).
To change the link functions $\eta_i=g_i(\theta_i)$ by means of which each EGPD parameter $\theta_i$ (here $\mu,\sigma$ or $\nu$) is indirectly optimised, or the default starting values for the parameters to fit, or to print these links, we use the commands:

\begin{lstlisting}[language=R]
EGPD1Family(mu.link="identity",mu.init=1, # log , inverse, own , ...
            sigma.link="log",sigma.init=3, nu.link="log",nu.init=0.5)
show.link(EGPD1Family)
\end{lstlisting}

It is possible to create an original link function "own", by gathering the expressions of the link function $\eta_i$, its inverse, the derivative of the latter w.r.t. parameter $\theta_i$ and its area of validity. For instance to create a log-link function shifted to the right for $\mu$:

\begin{lstlisting}[language=R]
# the link function defining the predictor eta  as a function of the current distribution parameter (referred to as mu), i.e. eta = g(mu)
own.linkfun <- function (mu) {
    .shift = 0.0001
    log (mu - .shift) 
}
#  the inverse of the link function as a function of the predictor eta, i.e.mu = g-1 (eta)
own.linkinv <- function (eta) {
    shift = 0.0001 
    thresh <- - log (.Machine$double.eps)
    eta <- pmin (thresh, pmax(eta, -thresh))
    exp (eta) + shift 
} 
# the first derivative of the inverse link with respect to eta
own.mu.eta <- function (eta) {
    shift = 0.0001  
    pmax (exp (eta), .Machine$double.eps) 
}
# the range in which values of eta are defined.
own.valideta <- function (eta) TRUE
EGPD1Family(mu.link="own")
\end{lstlisting}

For more details on link function definition, we refer to   \citet[][pag. 179]{stasinopoulos2017flexible}.

\section{Application to a hourly rainfall dataset in France}
\label{secApplRainfalls}

\subsection{Data}

We illustrate our add-on to package \texttt{gamlss} with the analysis of the hourly precipitations recorded over the north-west area of France from 2016 to 2018 both included. This database, containing precipitations as well as other meteorological variables whose values are registered every 6 minutes over the 3 years of interest for each of the 274 stations of the network (see Figure \ref{figSets}), is freely accessible  \citep[see][fro more details]{MeteoNet}.

Let us note $Y(\boldsymbol{x})$ the random variable for strictly positive hourly precipitations (dry and wet events are traditionally treated separately when it comes to rainfall simulation \citep{ailliot2015stochastic}), with covariates $\boldsymbol{x}=(x_l, x_L, x_t)$, where $x_l$ and $x_L$ represents the position in space (longitude, latitude) and $x_t$ a cyclic time index, namely the day or the month of the year. The work of \citet{naveau2016modeling} and our tests show that EGPD fitting is improved by censoring the data, so in practice we remove values $Y(\boldsymbol{x}) < 0.5$ (mm). Other thresholds ($\leq 0.2$ and $\leq 0.1$mm) have been tested but $Y(\boldsymbol{x}) < 0.5$ (mm) gives the  best results, in particular for the upper tails. Besides, as noted in \citet{evin2018stochastic,naveau2016modeling}, negative shape parameters $\xi$ are not expected for rainfall distributions at these time scales, so we use a link function constraining the parameter $\xi$ to positive values. Finally, as performed in \citet{naveau2016modeling}, we only keep one every three hourly records of precipitation so as to ensure independency between samples.

\subsection{Model fitting}

We fit different specifications of EGPD-based models, where EGPD takes form 1, 3 or 4, and compare them with equivalently specified Gamma-based models. The Gamma (GA) distribution is a classical choice that works well for modeling the bulk of precipitations at a given site \citep{katz1977precipitation,vrac2007stochastic,vlvcek2009daily}. However non-negligible deviations are observed when one is interested in capturing extreme rainfalls \citep{katz2002statistics}, typically underestimated due to the lightness of the GA tail. For a marginal model  \texttt{M=EGPD1,EGPD3,EGPD4,GA}, the different specifications read:
\begin{description}
	\item \texttt{M.0} where
$$	\eta_a(\boldsymbol{x})=\beta_0 , \qquad a=\mu,\sigma,\nu,\tau,$$
i.e.	 the parameters $\theta(\boldsymbol{x})$ do not depend on the covariates ;
	\item  \texttt{M.t} where 
			$$
	\eta_a(\boldsymbol{x})=\beta_0+ s_t (x_t) , \qquad a=\mu,\sigma,\nu,\tau,
	$$
		uses cyclic cubic splines $s_t$ to model the nonlinear dependency in time while considering no variation in space ;
	\item \texttt{M.tnomu} where 
			$$
	\eta_a(\boldsymbol{x})=\beta_0+ s_t (x_t) , \qquad a=\sigma,\nu,\tau,
	$$ does the same as before but considers as constant the parameter $\mu$ ;
	\item \texttt{M.st} where 
			$$
	\eta_a(\boldsymbol{x})=\beta_0+ TP(x_l,x_L)+ s_t (x_t), \qquad a=\mu,\sigma,\nu,\tau,
	$$
	uses smooth surface fitting (thin-plate splines) to model the dependency in space as well as cyclic cubic splines for the dependency in time ;
	\item \texttt{M.st2mu} where 
			$$
	\eta_a(\boldsymbol{x})=\beta_0+ TP(x_l,x_L)+ s_t (x_t), \qquad a=\mu,\sigma,
	$$
	does the same as before but considers constant in space the parameters that are not $\mu$ or $\sigma$ ;
	\item \texttt{M.st2nomu} where 
			$$
	\eta_a(\boldsymbol{x})=\beta_0+ TP(x_l,x_L)+ s_t (x_t), \qquad a=\sigma,\nu,
	$$
	does the same as before but considers constant in space the parameters $\mu$ and $\tau$ ;
	\item \texttt{M.st3nomu} where 
			$$
	\eta_a(\boldsymbol{x})=\beta_0+ TP(x_l,x_L)+ s_t (x_t), \qquad a=\sigma,\nu,\tau,
	$$
	does the same as before but applies to \texttt{M=EGPD4} only and considers parameter $\mu$ as a constant over space.
\end{description}

\begin{figure}[h!]
	\begin{center}
		\includegraphics[width=0.5\columnwidth]{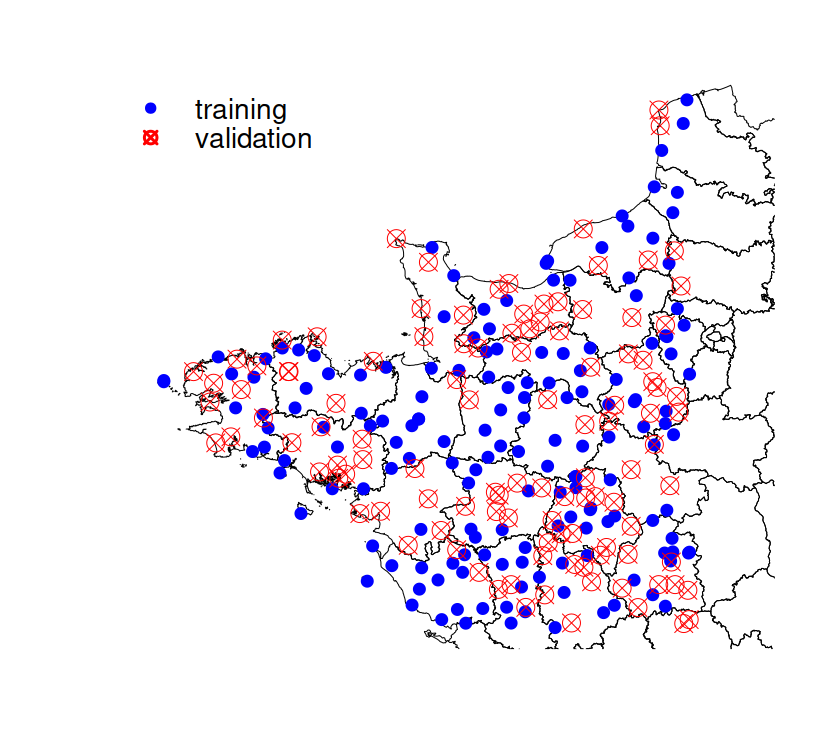}
		\caption{French stations from the MeteoNet public dataset used for our experiments, split into training and validation sets.
			{\label{figSets}}
		}
	\end{center}
\end{figure}

These variations within each model class will allow us to assess whether the smooth regression of distribution parameters over covariates is justified or not. 
Below we reproduce the code to fit \texttt{megpd1.0}, \texttt{megpd1.tnomu} and \texttt{mga.st} on the training set (\texttt{training}), that contains 60 \% of the stations, as reported on Figure \ref{figSets}. The dataset consisting of the remaining stations form the validation set (\texttt{validation}). The \texttt{gamlss()} function reads:

\begin{lstlisting}[language=R]
# Notations: lat, lon and cyc represent, respectively, covariates xl, xL and xt, while y is given by precip
# Set algorithm control parameters:
con <- gamlss.control (n.cyc = 200, mu.step = 0.01, sigma.step = 0.01, nu.step = 0.01,tau.step = 0.01, autostep=TRUE)
megpd1.0 <- gamlss(precip ~ 1,
       data = training, 
                  family = EGPD1Family(mu.link = "own"),
                  control = con,
                  method=CG())         
megpd1.tnomu <- gamlss(precip ~ 1,
                  sigma.fo =~ pbc(cyc),
                  nu.fo =~ pbc(cyc),
                  data = training, 
                  family = EGPD1Family(mu.link = "own"),
                  control = con,
                  method=CG()) 
                      
# Load library to interface with the mgcv package and get ga() for multidimensional tensor spline smoothers:
library(gamlss.add)
fmla_ga <- list(~s(lon,lat,bs='tp',k=30)+s(cyc,bs="cc",k = 50),
                ~s(lon,lat,bs='tp',k=30)+s(cyc,bs="cc",k = 50))
GA.st <- gamlss(precip ~ga(~s(lon,lat,bs='tp',k=30)+s(cyc,bs="cc",k = 50),                                          method="REML"), 
                      ~ga(~s(lon,lat,bs='tp',k=30)+s(cyc,bs="cc",k = 50), 
                      method="REML"),
                      ~ga(~s(lon,lat,bs='tp',k=30)+s(cyc,bs="cc",k = 50), 
                      method="REML"),
                      data = training, 
                      family = GA,
                      control = con,
                      method=CG())
\end{lstlisting}

\subsection{Model assessment}

\begin{figure}[h!]
	\begin{center}
		\includegraphics[width=0.85\columnwidth]{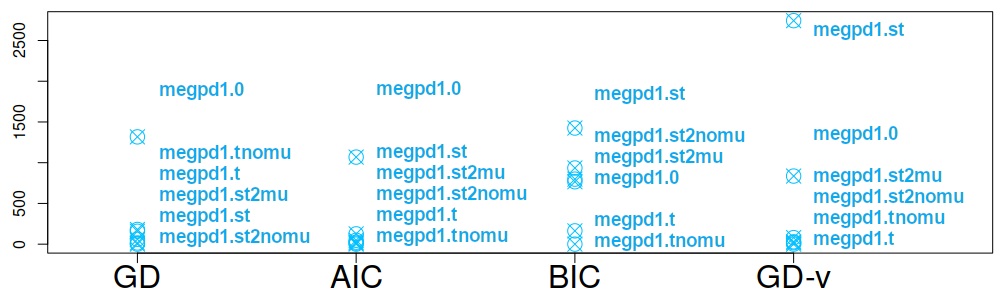}
		\includegraphics[width=0.85\columnwidth]{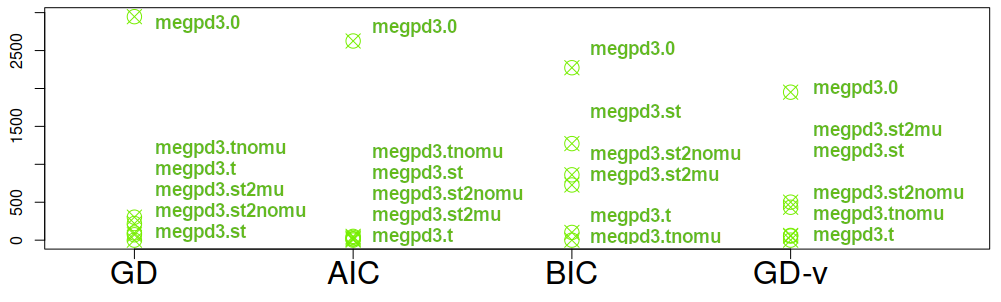}
		\includegraphics[width=0.85\columnwidth]{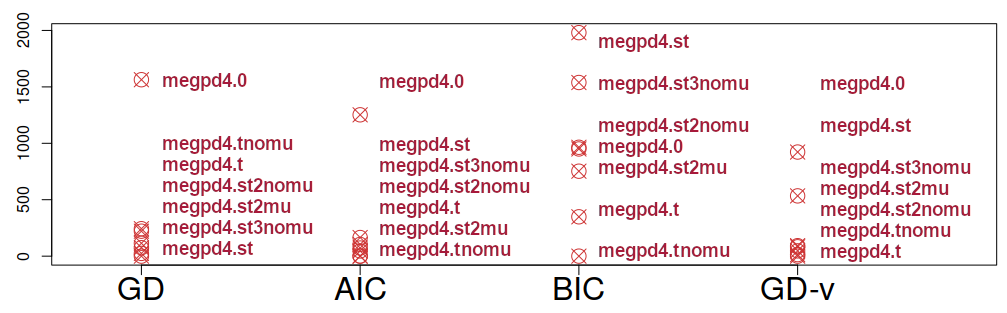}
		\includegraphics[width=0.85\columnwidth]{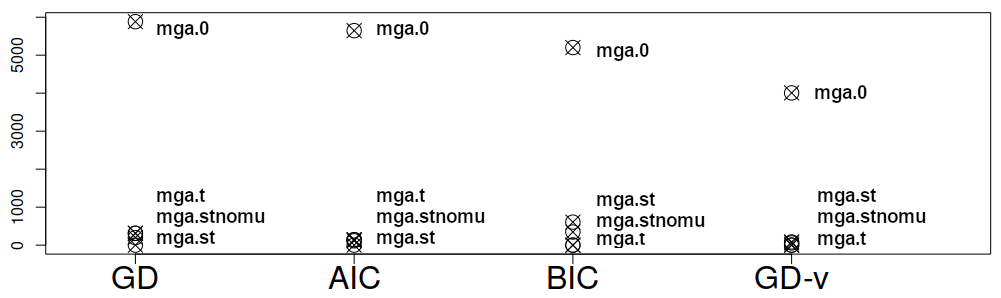}
		\caption{Values above the global minimum of each information criterion associated with fitted models: global deviance (GD), AIC, BIC on fitting set, and GD on validation set (GD-v). From top to bottom, we compare the variations of EGPD1, EGPD3, EGPD4 and Gamma-based models. The model variations are indicated in full letters on the right side of the points, in an order following their relative ranking.
			{\label{figInfCriteria}}%
		}
	\end{center}
\end{figure}

In order to evaluate the goodness of fit on the training set, we use summary statistics: Global Deviance (GD), Akaike Information Criterion (AIC) and Bayesian Information Criterion (BIC).
The \texttt{GAIC(modelName,k=a)} function allows to compute GD (minus twice the fitted log-likelihood, $a=0$), the AIC ($a=2$) and the BIC ($a=\log(\mbox{number of observations})$) for each fitted model \texttt{modelName} and to compare them as performed on Figure \ref{figInfCriteria}. Besides, Table \ref{table1} indicates the effective degree of freedom (df) computed for each model. The function call is reproduced below for the models of class EGPD1.

\begin{lstlisting}[language=R]
GAIC_BIC_m1 <-GAIC(megpd1.0,megpd1.t,megpd1.tnomu,megpd1.st,megpd1.st2mu,megpd1.st2nomu,k=log(length(training$precip))
\end{lstlisting}

\begin{table}[h]
	\begin{center}
		\begin{tabular}{||l c||} 
			\hline
			Model & \multicolumn{1}{c||}{df} \\ [0.5ex] 
			\hline\hline
			\texttt{megpd4.st}  & 242.0 \\
			\hline
			\texttt{megpd4.st3nomu}  & 199.3 \\
			\hline
			\texttt{megpd4.st2nomu}  & 138.3 \\
			\hline
			\texttt{megpd4.st2mu}  & 122.5 \\
			\hline
			\texttt{megpd4.t}  & 72.0 \\
			\hline
			\texttt{megpd4.tnomu} & 38.0 \\ 
			\hline
			\texttt{megpd4.0} & 4 \\
			\hline
			\texttt{megpd1.st}  & 183.0 \\
			\hline
			\texttt{megpd1.st2nomu}  & 139.3 \\
			\hline
			\texttt{megpd1.st2mu}  & 121.3 \\
			\hline
			\texttt{megpd1.t}  & 54.0 \\
			\hline
			\texttt{megpd1.tnomu}  & 37.0 \\
			\hline
			\texttt{megpd1.0}  & 3 \\
			\hline
			\texttt{megpd3.st}  & 181.6 \\
			\hline
			\texttt{megpd3.st2nomu}  & 137.3 \\
			\hline
			\texttt{megpd3.st2mu}  & 122.9 \\
			\hline
			\texttt{megpd3.t}  & 54.0 \\
			\hline
			\texttt{megpd3.tnomu}  & 37.0 \\
			\hline
			\texttt{megpd3.0}  & 3 \\
			\hline
			\texttt{mga.st}  & 120.2 \\
			\hline
			\texttt{mga.stnomu}  & 78.0 \\
			\hline
			\texttt{mga.t}  & 36.0 \\
			\hline
			\texttt{mga.0}  & 2 \\ [1ex] 
			\hline
		\end{tabular}
	\end{center}
	\caption{\label{table1} Effective degree of freedom (df) for the fitted models.}
\end{table}

From Figure \ref{figInfCriteria}, we can see that, as could be expected, for each classes of models (where EGPD1-based models represent one class, EGPD3-based ones another, etc) the GD ranks first models with the full space-time description \texttt{M.st}. All other variations are rather close in fitting performance, at the exception of the constant-parameters one \texttt{M.0}, that is largely behind. This shows the clear interest of allowing a smooth variation of distribution parameters as a function of the chosen covariates. However, when it comes to modeling efficiency, where the trade-off between number of parameters and performances is taken into account (through the AIC and BIC), the full space-time model \texttt{M.st} becomes much less desirable, falling just  above and occasionally behind the constant parameter models \texttt{M.0}. This is all the more through that we consider the BIC indice, that penalizes more the number of parameters than the AIC, and a model with a large number of parameters (EGPD4 versus GA). The best trade-off, according to the BIC, are the time-only full model \texttt{M.t} for the GA class and the time-only without the shape parameter $\mu$ otherwise. This may also be an indication that the seasonal variation of distribution parameters is more important than their spatial variation over the area of interest. 

The ranking of model variations within a same model class allows us to draw conclusions w.r.t. the advantages or challenges of each class. Thus, for EGPD4 the GD ranks model variations following decreasing complexity: \texttt{megpd4.st} followed by \texttt{megpd4.st3nomu}, etc. We can deduce that the estimation of $\mu$ is not that sensitive when it comes to the fitting set (at least with enough data). On the contrary, \texttt{megpd1.st2nomu} is preferred to \texttt{megpd1.st} and \texttt{megpd1.st2mu} which can be interpreted as $\mu$ being a rather sensitive parameter to estimate for EGPD1. Considering it constant in space consequently improves the global fitting. 

Performances on the validation set are obtained by means of the \texttt{getTGD()} and \texttt{TGD()} functions, providing the validation global deviance (GD-v), that is the GD evaluated using predictive values for the parameters at the validation sample.
\begin{lstlisting}[language=R]
	gg.megpd4.st <- getTGD(megpd4.st,newdata=validation)
	gg.megpd4.0 <- getTGD(megpd4.0,newdata=validation)
	...
	TGD(gg.megpd4.0,gg.megpd4.t,gg.megpd4.tnomu,gg.megpd4.st,gg.megpd4.st2mu,gg.megpd4.st2nomu,gg.megpd4.st3nomu)
\end{lstlisting}
Thus, when it comes to extrapolating parameters for new covariate values, the GD-v ranking displayed in Figure \ref{figInfCriteria} reveals that models considering only the dependence in time (\texttt{M.t}, \texttt{M.tnomu}) are systematically preferred, followed by space-time models not considering $\mu$ (\texttt{M.st2nomu}), before the full space-time models (\texttt{M.st}). This supports the fact that the rainfall distributions are relatively close over the studied geographical area, compared to their time evolution that shows stronger variations. Consequently, given the limited size of the dataset, keeping low the number of parameters to regress on covariates (by considering the seasonality only) and in particular avoiding in this category the shape parameter $\mu$, makes estimations more robust. This agrees with observations from \citet{evin2018stochastic}, who noted that fitting the shape parameter is the most challenging part of (extended)-GPD modeling. They concluded that mono-site fitting is not robust, leading e.g. to negative shape parameters that do not match observations on rainfall intensities. To circumvent this issue, the authors use constant parameters within regions of influence defined by homogeneity tests.

\subsubsection{Model verification and comparison}
\begin{figure}
	\begin{center}
		\includegraphics[width=0.48\columnwidth]{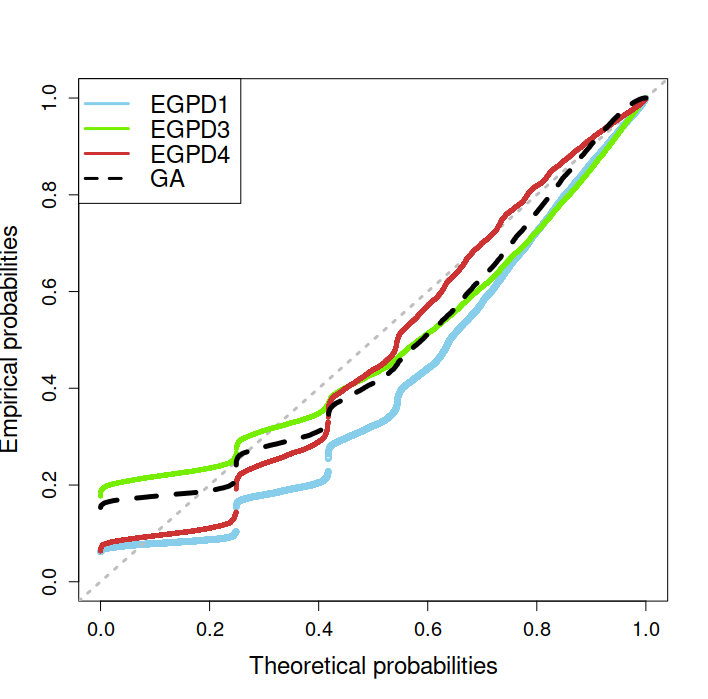}
		\includegraphics[width=0.48\columnwidth]{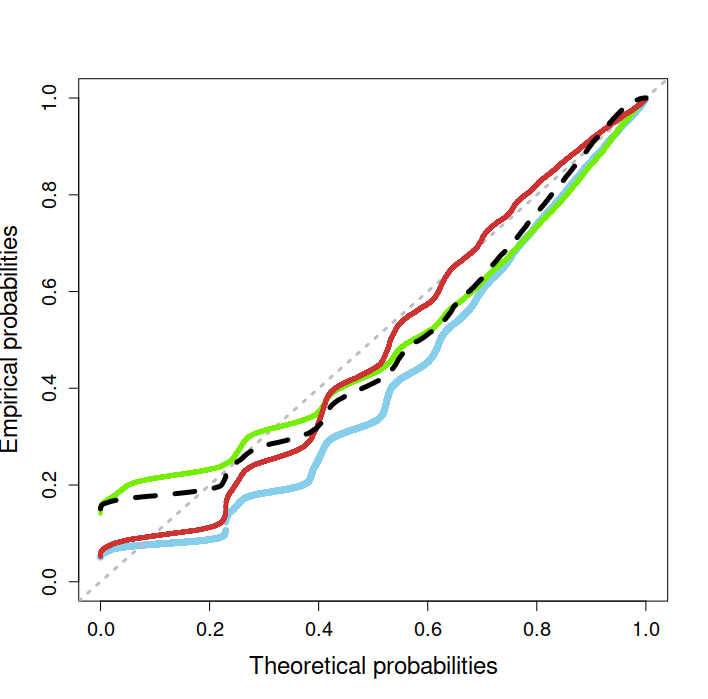}
		\includegraphics[width=0.48\columnwidth]{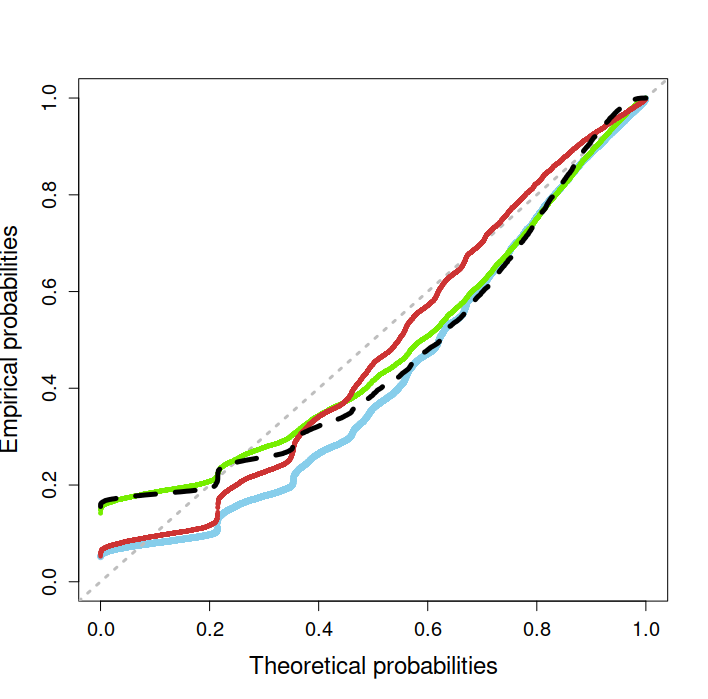}
		\includegraphics[width=0.48\columnwidth]{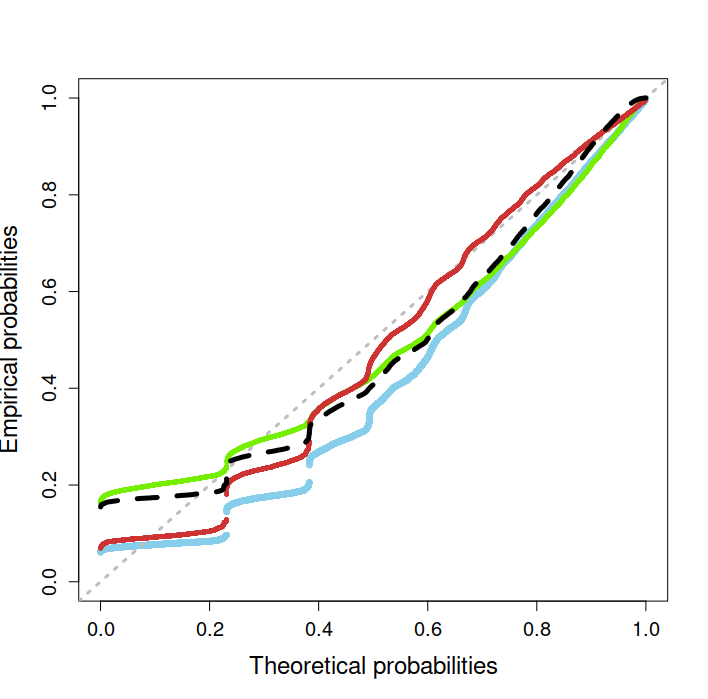}
		\caption{Assessment of the quality of each model (EGPD1, EGPD3, EGPD4 and GA) for all stations from the verification set in (from left to right and top to bottom) winter, spring, summer and autumn. The closer to the diagonal, the better the underlying model distribution. See text for theoretical development. We compare the \texttt{st2nomu} version of each model class (that is where parameters $\mu,\tau$ are constant in space). Theoretical probabilities $\leq 0.4$ is overall equivalent to $y \leq 1$ mm where $y$ designates the hourly rainfall amount.
			{\label{figUplots}}%
		}
	\end{center}
\end{figure}

Now that it is admitted that regressing distributional parameters over time and possibly space is justified for the case study at hand, we want to assess how good is the modeling and compare model classes at hand.
A way to verify whether the fitted models actually represents well the empirical distribution is to draw so-called P-P plots. 
A simple result in probability states if $Y(\mathbf{x}) \sim F(\cdot;\theta(\mathbf{x})$ then $U(\mathbf{x}):=F(Y(\mathbf{x});\theta(\mathbf{x}))$
is uniformly distributed on the unit interval. 

Therefore we can consider the  residuals $u(\mathbf{x}) = F(y(\mathbf{x}),\widehat{\theta}(\mathbf{x}))$ where $\widehat{\theta}(\mathbf{x})$ are the parameters
 of the fitted model and, after having ordered them,  compare them with  $\Big( \frac{1}{n+1}, ..., \frac{n}{n+1} \Big)$. Here $n$ is the number of observations $y(\mathbf{x})$ at hand. The closer points are to the diagonal, the better the model fit. Results are broken down at each season for model variations \texttt{M.st2nomu} on the verification set in Figure \ref{figUplots}. 
They show that at all seasons, EGPD4 outperforms the other models. This is all the more true that we consider not too small rainfalls, namely hourly amounts $x \geq 1$ mm. Below, we keep wiggly artefacts of the discrete sampling scale of observations. Indeed, rainfall amounts are sampled with a precision of $0.2$ mm, which makes the distribution of small values more discrete than continuous. Behind, come models GA, EGPD3 and finally EGPD1. EGPD3 is often better than GA for small hourly rain amounts (below around $2$ mm) however it may be outperformed for larger ones, especially in winter. EGPD1 performs as EGPD3 for intermediate and large amounts of rain, but is significantly behind all other models for small amounts.

\subsubsection{Visualisation of the fitted functions}

\begin{figure}
	\begin{center}
		\includegraphics[width=0.75\columnwidth]{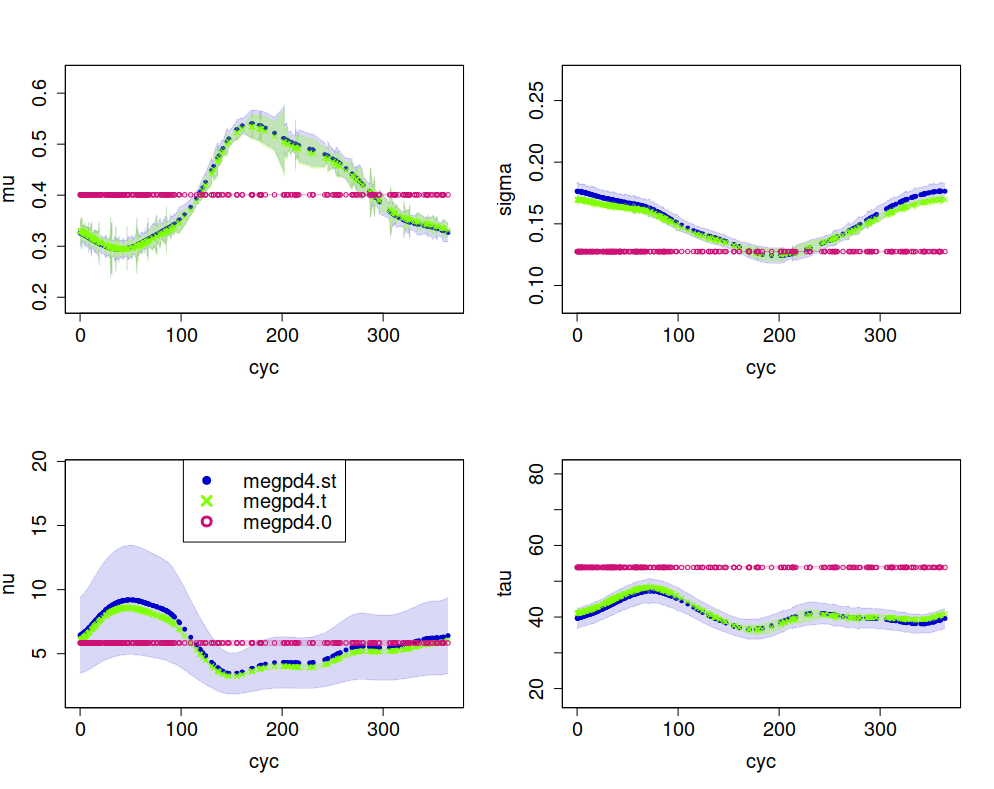}
		\includegraphics[width=0.75\columnwidth]{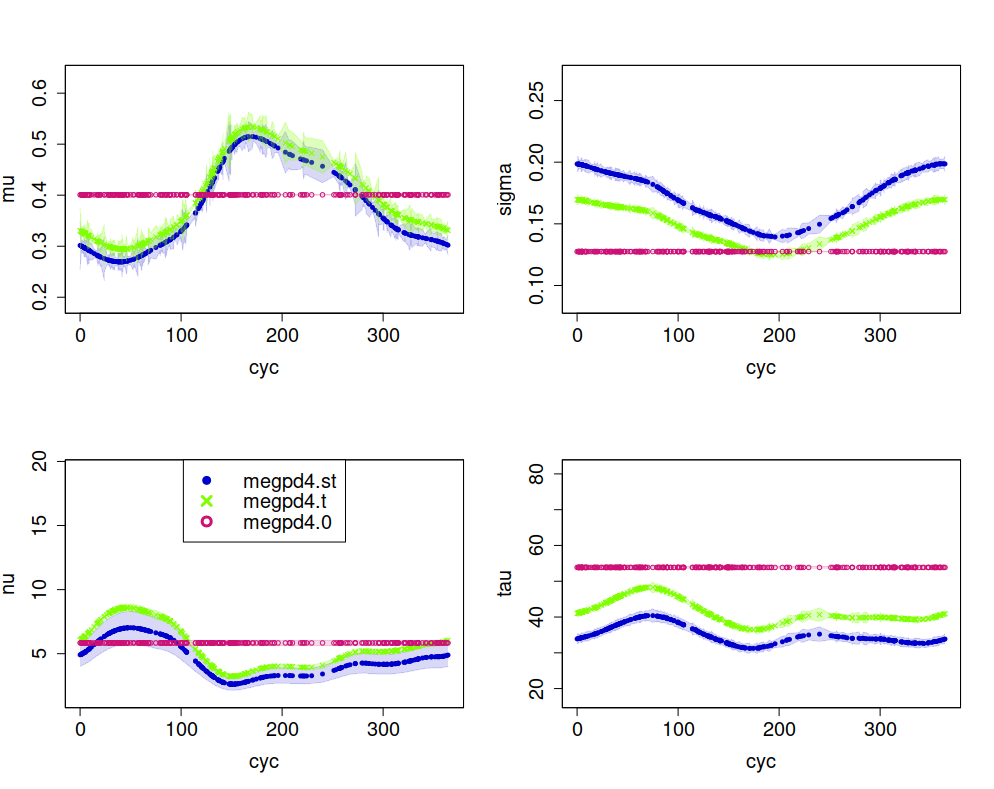}
		\caption{EGPD4 coefficients fitted for the models \texttt{megpd4.0} (constant parameters), \texttt{megpd4.t} (smooth time dependence only), and \texttt{megpd4.st} (full smooth space time dependence) in two stations: (top two rows) on the north-west coast of Brittany ($x_L=48.7$ $x_l=-4.33$) and (bottom two rows) in the south-east countryside of the studied area ($x_L=46.7$ $x_l=1.25$). Coefficients are fitted against the covariate $cyc$, namely time of the year. The shadowed areas correspond to the associated standard errors.
			{\label{figParEvol}}%
		}
	\end{center}
\end{figure}

Figure \ref{figParEvol} represents the fitted functions for the variations \texttt{megpd4.0} (constant parameters), \texttt{megpd4.t} (smooth time dependence only), and \texttt{megpd4.st} (full smooth space time dependence) of the EGPD4 model in two \textit{a priori} rather different stations when it comes to the rainfall regime: the north-west coast of Brittany ($x_L=48.7$ $x_l=-4.33$) and the south-east countryside of our area of interest ($x_L=46.7$ $x_l=1.25$).
 
The functions \texttt{predict()} or \texttt{predictAll()} are used to provide fitted values of the functions of the covariates. Standard errors can be extracted for fitted values by means of the argument \texttt{se.fit=TRUE}. However this option is not supported by \texttt{gamlss} for parameter estimates on new covariate values at the moment of writing this paper. 
\begin{lstlisting}[language=R]
# Fitted parameter values:
muFit <- predict(megpd4.st,what="mu",type="response")

# Estimates for new covariate values, gathered in the database "validation":
all <- predictAll(megpd4.st,data=training,newdata=validation)
muVal <- all$mu
sigmaVal <- all$sigma
nuVal <- all$nu

muFit <- predict(megpd4.st,what="mu",type="response",data=training,se.fit=TRUE)
StandardErrorMu <- muFit$se.fit
\end{lstlisting}
We observe rather similar temporal behaviors of the parameters estimated in both stations, highlighting the relative spatial homogeneity of distributions compared to their seasonal dependence. The standard errors decrease with the simplicity of the model ; they are generally negligible for for parameters $\sigma$ and $\tau$ but can become significant for parameters $\nu$ and $\mu$ when we are dealing with space-time models. Beyond that, the estimations for the three different versions of the EGPD4-based model are consistent. 
We note (second station) that a shift between \texttt{megpd4.t} and \texttt{megpd4.st}'s estimates of parameter $\sigma$ is compensated by an opposite shift between the associated estimates of parameter $\tau$.
From our observations and experience on EGPD fitting, a given EGPD-based distribution is matched by several local optima (i.e. parameter combinations). Along this "Pareto front", $\nu$ is increasing for EGPD1 when $\sigma$ decreases. Similarly  $\tau$ is increasing for EGPD4 when $\sigma$ decreases. 

\begin{figure}[h!]
	\begin{center}
		\includegraphics[width=\columnwidth]{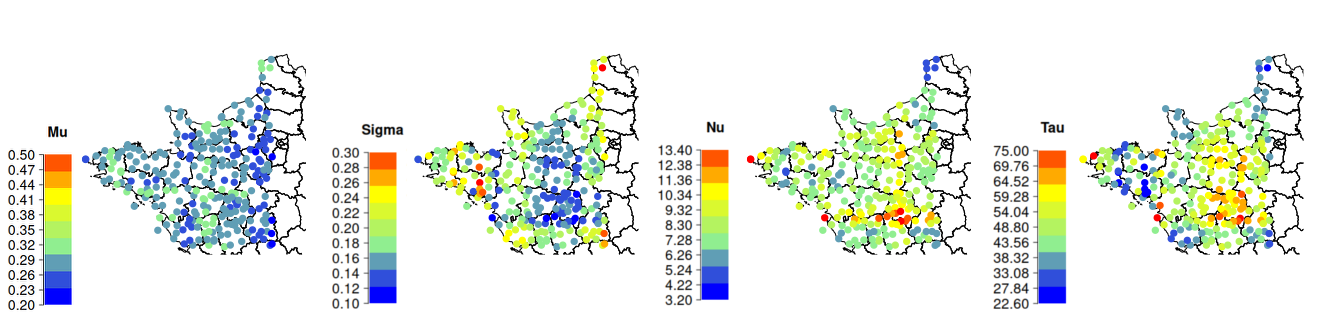}
		\includegraphics[width=\columnwidth]{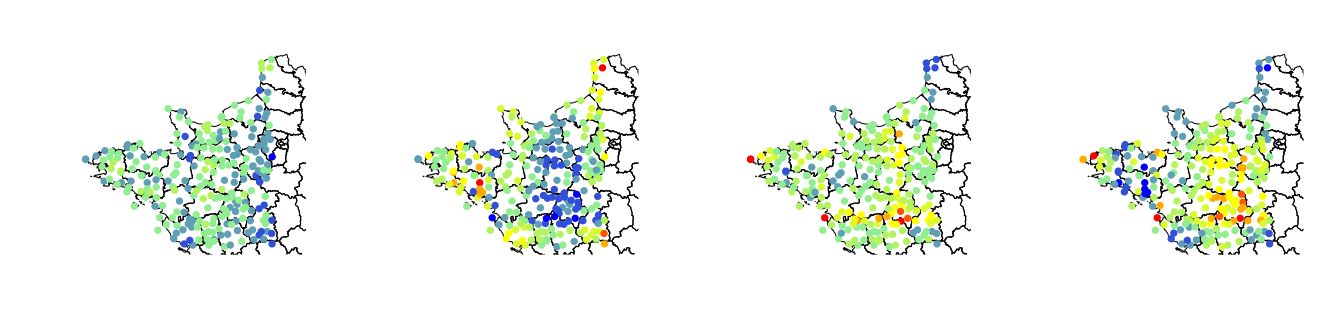}
		\includegraphics[width=\columnwidth]{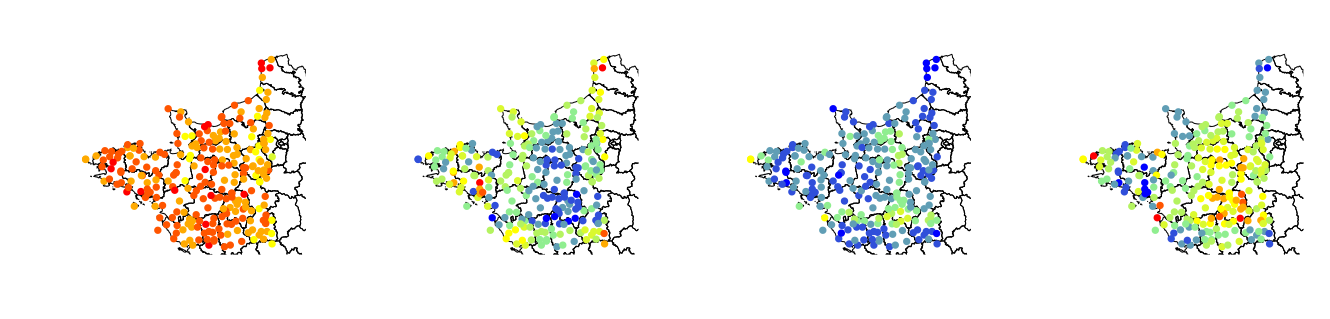}
		\caption{Median parameter maps for the full EGPD4 model \texttt{megpd4.st} for the months of December (top row), April (middle row) and August (last row).
			{\label{figParMaps}}%
		}
	\end{center}
\end{figure}

Finally, Figure \ref{figParMaps} represents the spatial distribution of the median of the EGPD4 parameters estimated from \texttt{megpd4.st} at three different times of the year: december, april and august. We indeed observe spatial variations of the parameter values, in addition to the monthly changes already discussed. $\mu$ the most spatially homogeneous parameter, which goes in line with previous observations that favored spatially constant $\mu$-models over non-constant ones.
We also note that the $\tau$ map follows an opposite trend w.r.t. $\sigma$'s map: the larger $\tau$, the smaller $\sigma$, which follows the general idea of parameter Pareto front developed earlier.

\subsubsection{Goodness-of-fit w.r.t. the upper tail}
\begin{figure}
	\begin{center}
		\includegraphics[width=0.3\columnwidth]{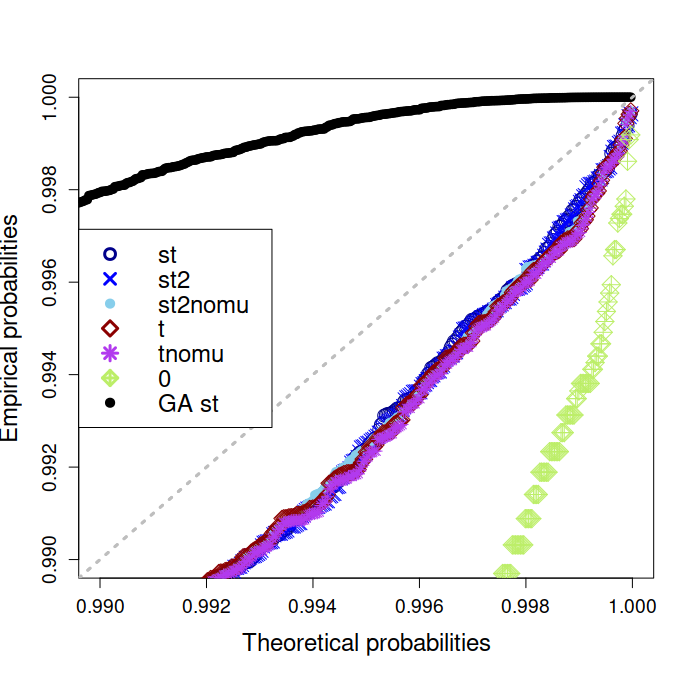}
		\includegraphics[width=0.3\columnwidth]{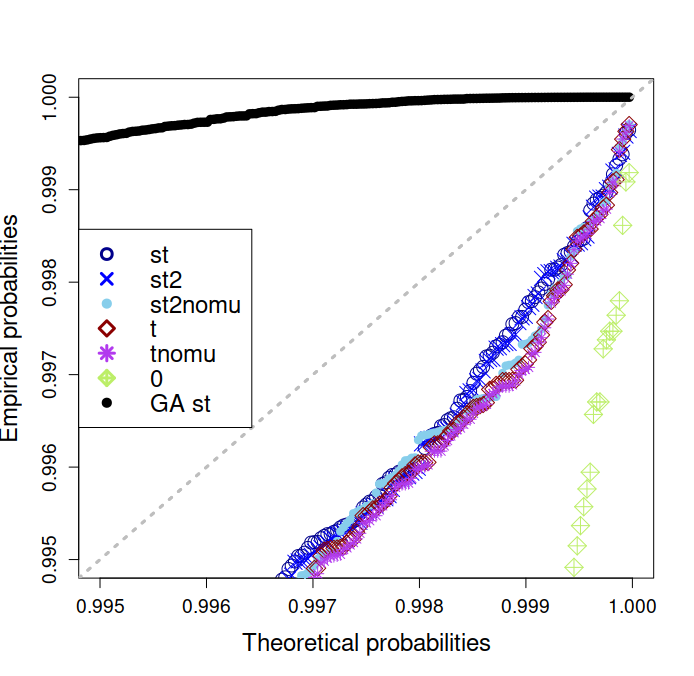} \\
		\includegraphics[width=0.3\columnwidth]{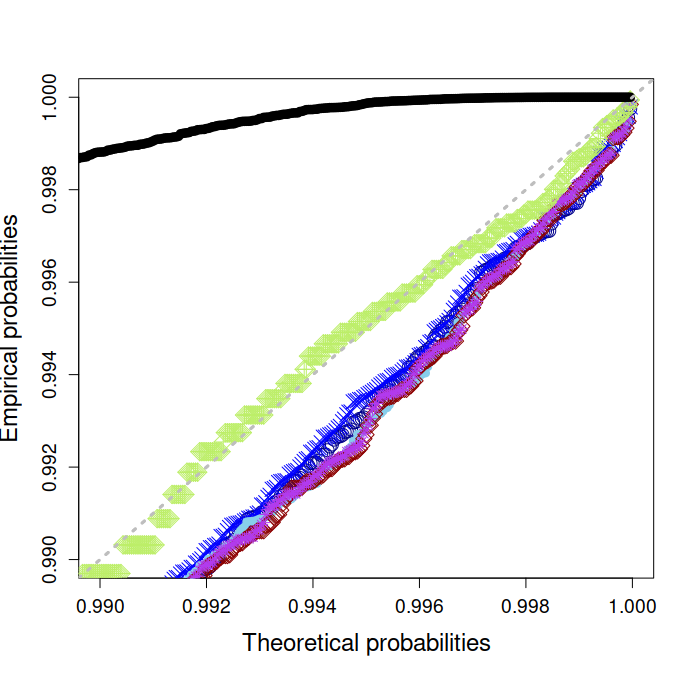}
		\includegraphics[width=0.3\columnwidth]{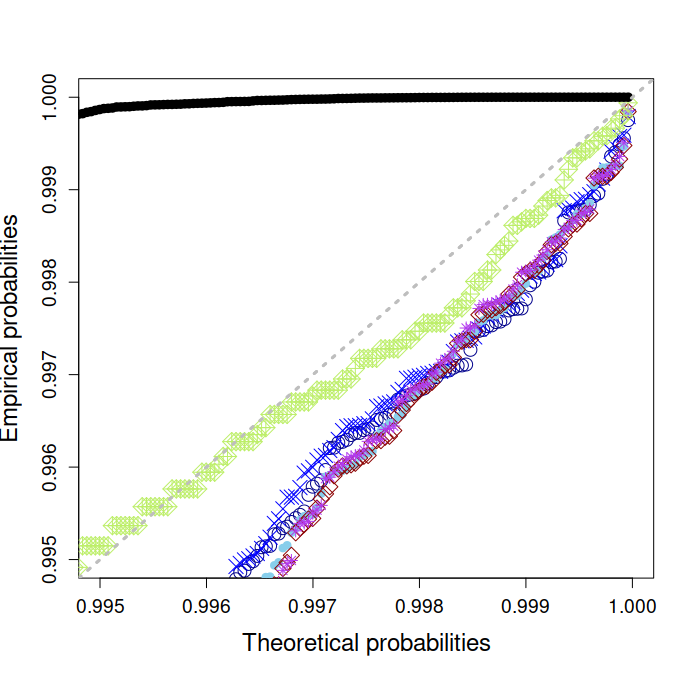} \\
		\includegraphics[width=0.3\columnwidth]{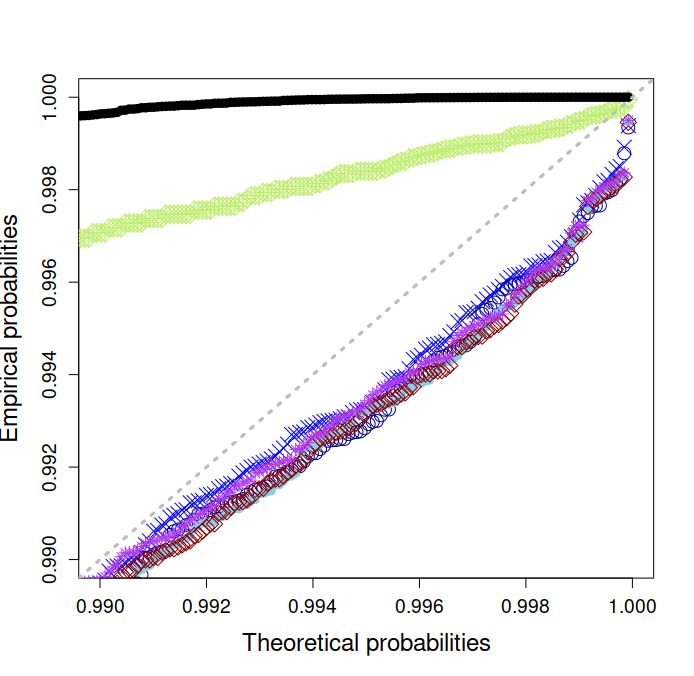}
		\includegraphics[width=0.3\columnwidth]{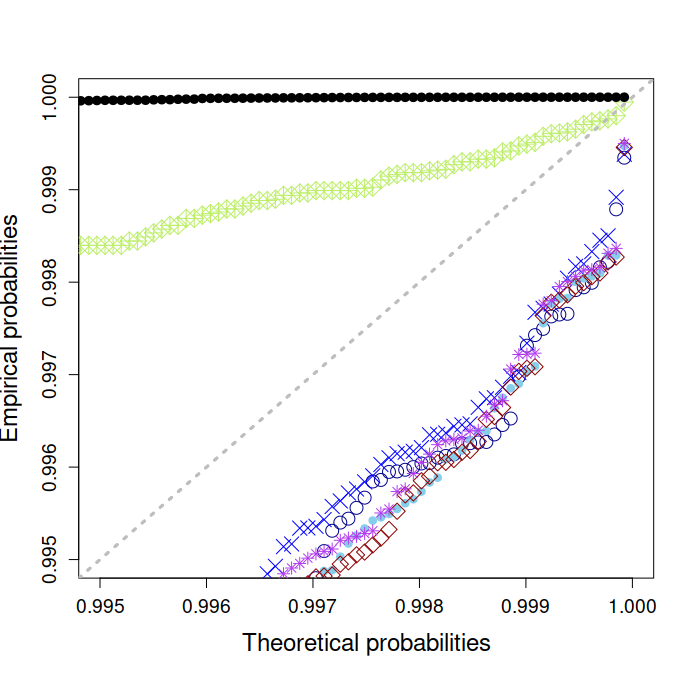} \\
		\includegraphics[width=0.3\columnwidth]{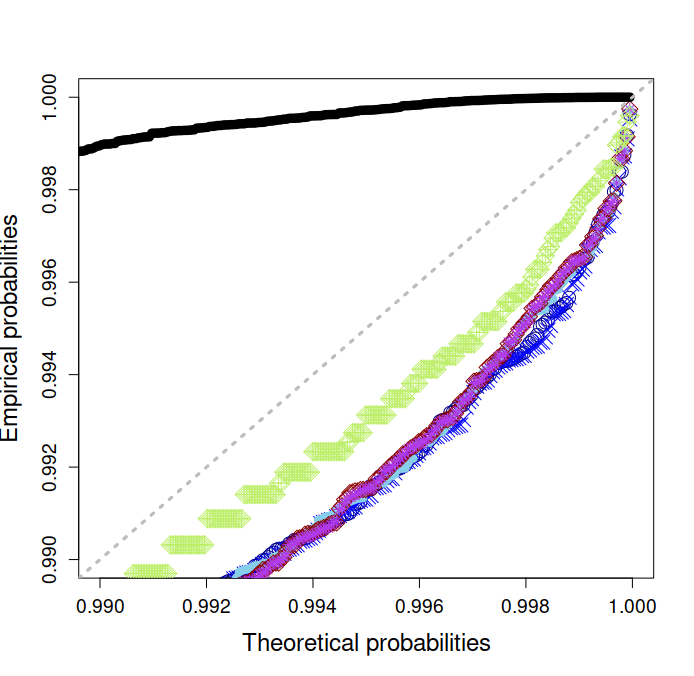}
		\includegraphics[width=0.3\columnwidth]{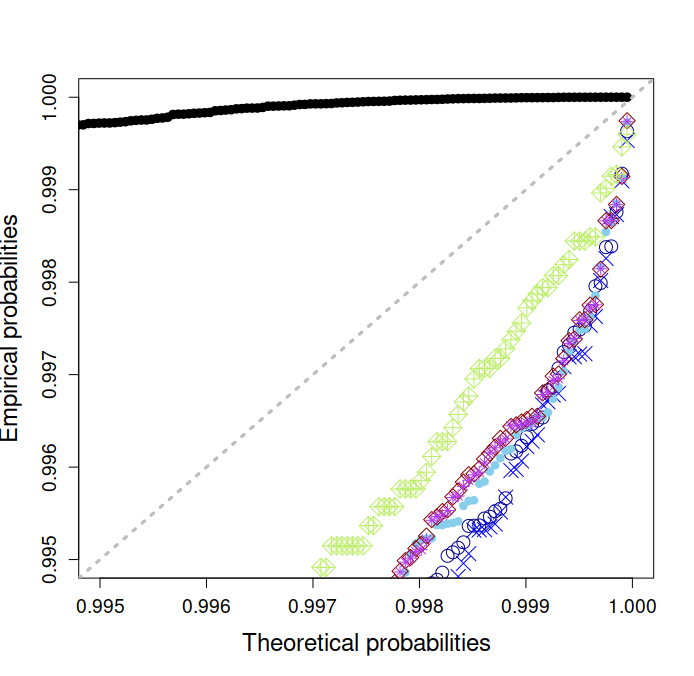}
		\caption{Assessment of the quality of model EGPD4's variations and GA for all stations from the validation set in (from top to bottom) winter, spring, summer and autumn. The closer to the diagonal, the better the underlying model distribution, see Figure \ref{figUplots}. We focus on extremes, with theoretical probabilities $\geq 0.99$ (left column) equivalent to hourly rainfall amounts above $11$ mm for EGPD4 models and above $7$ mm for GA. Theoretical probabilities $\geq 0.995$ (right column) are equivalent to rain amounts above $15$ mm for EGPD4 models and above $7.6$ mm for GA.
		{\label{figUplotext}}%
		}
	\end{center}
\end{figure}
We finish the presentation of the results by comparing in Figure \ref{figUplotext} the EGPD4 model to a GA fit, for all stations of the validation set and for each season. We  want to check the added value of the EGPD4 model over the GA when it comes to extremes, as well as to assess the best modeling option for the EGPD4 shape parameter in order to ensure a correct representation of extremes.
The series of P-P plots presented on Figure \ref{figUplotext} focus on the upper tail of the distribution (above the quantiles 0.99 and 0.995 respectively) and allow us to do that. We first note that whatever the season, EGPD4 performs significantly better than GA when it comes to extremes. Most often, the constant parameter model \texttt{megpd4.0} is enough to outperform the GA, and sometimes is clearly the best option as in spring and autumn. In winter and summer, it is not enough to capture correctly the upper tail of the empirical distribution. Making EGPD4 parameters dependent on time only is sufficient to obtain correct upper tail representations, whether $\mu$ is considered constant (\texttt{megpd4.tnomu}) or not (\texttt{megpd4.t}). The difference between the last two models is very small however using the constant shape parameter option can slightly improve results, especially for larger quantiles (0.995 and above versus 0.99). Adding spatial dependence to the temporal one improves even more the accuracy of the EGPD4 upper tail modeling. Making spatially dependent only two parameters usually performs better. However $\mu$ needs to be included in these two parameters (\texttt{megpd4.st2} versus \texttt{megpd4.st2nomu}). 
This highlights the fact that a smooth space-time estimation of the shape parameter $\mu$ is important to capture correctly the non-stationary behavior of the upper tail  of the rainfall distributions. However, due to limited data and numerical sensitivity of that estimation, it may be preferable to only limit the smooth spatial dependence to two parameters instead of four, including $\mu$, in order to make it more robust for extrapolation.

\section{Conclusions}
\label{secConclu}

This article introduced an add-on to the R-package \texttt{gamlss}, to model non-stationary fields whose margins are EGPD distributed. A number of R packages target the modeling of heavy-tailed distributions. Yet, although some allow distributional regression models, for which distribution parameters can be modeled as additive functions of covariates, none of these packages address the extended-GPD distribution nor another form of unique and closed-form formulation for heavy-tailed distributions that models small, intermediate and large values in a synthetic manner. This add-on implements the EGPD family in a generic way in the \texttt{gamlss} package, allowing to test any new parametric form, up to 4 parameters so far, satisfying the requirement of the EGPD family. 
Applications include space-time modeling of rainfall marginal distributions. The example developed in this paper shows 1) that modeling hourly rainfall amounts is significantly improved when distribution parameters are non-stationary in space and time; and that 2) the EGPD class clearly improves the modeling of intermediate values and upper tail compared to a classical Gamma option. This illustrates the benefits of our add-on with respect to existing GAM forms for modeling non-stationary margins (typically the Gamma model for our application). We compare three parametric EGPD models and discuss the modeling strategies to best capture non stationary upper tails with limited datasets. This case study also reveals the added-value of EGPD4 (and to a less extent EGPD3) over EGPD1 to correctly model hourly rainfall distributions in a space-time context, while previous works \cite{naveau2016modeling,evin2018stochastic,bertolacci2018comparison} restricted themselves to EGPD1 as the best option.


\section*{Software and data availability}

The add-on to the R-package \texttt{gamlss} \citep{gamlssP} presented here for the generic implementation of EGPD consists in the script \texttt{GenericEGPDg.R}, available at \url{https://github.com/noemielc/egpd4gamlss}. On this same webpage, supplementary material consisting in a reproducible tutorial with code is presented.

The data used in this paper is freely accessible on the MeteoNet French database \citep{MeteoNet}: \url{https://meteofrance.github.io/meteonet/english/data/summary/}

\section*{Acknowledgement}

Scientific activity was performed as part of the research program Venezia2021, coordinated by CORILA, with the contribution of the Provveditorato for the Public Works of Veneto, Trentino Alto Adige and Friuli Venezia Giulia (Italy).
The authors thank Nicolas Berthier for helpful R suggestions.


\bibliographystyle{elsarticle-harv.bst}
{
	\setlength\baselineskip{.96\baselineskip}%
	\bibliography{egpdbib}}

\end{document}